\documentstyle[twocolumn,aps,floats,psfig]{revtex}
\begin{document}
\author{Mauricio Cataldo\thanks{%
E-mail address: mcataldo@alihuen.ciencias.ubiobio.cl}}
\address{Departamento de F\'\i sica, Facultad de Ciencias,
Universidad del B\'{\i}o-B\'{\i}o, \\
Avda. Collao 1202, Casilla 5-C, Concepci\'on, Chile and \\
Departamento de F\'\i sica, Facultad de Ciencia,
Universidad de Santiago de Chile, \\
Avda. Ecuador 3493, Casilla 307, Santiago, Chile}
\author{Alberto Garc\'\i a\thanks{%
E-mail address: aagarcia@fis.cinvestav.mx}}
\address{Departamento de F\'\i sica, Centro de
Investigaci\'on y de Estudios Avanzados del IPN, \\ Apartado Postal
14--740, C.P. 07000, M\'exico, D.F. M\'exico, and \\ Departamento
de F\'\i sica, Facultad de Ciencia, Universidad de Santiago de
Chile,
\\ Avda. Ecuador 3493, Casilla 307, Santiago, Chile}
\title{Regular (2+1)-dimensional black holes within non-linear
Electrodynamics.}
\maketitle
\begin{abstract}
{\bf Abstract:}{(2+1)-regular static black hole solutions with a
nonlinear electric field are derived. The source to the Einstein
equations is an energy momentum tensor of nonlinear
electrodynamics, which satisfies the weak energy conditions and in
the weak field limit becomes the (2+1)-Maxwell field tensor. The
derived class of solutions is regular; the metric, curvature
invariants and electric field are regular everywhere. The metric
becomes, for a vanishing parameter, the (2+1)-static charged BTZ
solution. A general procedure to derive solutions for the static
BTZ (2+1)-spacetime, for any nonlinear Lagrangian depending on the
electric field is formulated; for relevant electric fields one
requires the fulfillment of the weak energy conditions.\\}


\end{abstract}

\smallskip\

In general relativity the literature on regular black hole solutions is
rather scarce~\cite{Borde,Garcia}. In (3+1)-gravity it is well known that
electrovacuum asymptotically flat metrics endowed with timelike and
spacelike symmetries do not allow for the existence of regular black hole
solutions. Nevertheless, in the vacuum plus cosmological constant $\Lambda$
case, the de-Sitter solution~\cite{Kottler} with positive cosmological
constant is known to be a regular non-asymptotically flat solution (the
scalar curvature is equal to $4 \Lambda$ and all the invariants of the
conformal Weyl tensor are zero.) In order to be able to derive regular
(black hole) gravitational--nonlinear electromagnetic fields one has to
enlarge the class of electrodynamics to nonlinear ones~\cite{Garcia}. On the
other hand in (2+1)-gravity, which is being intensively studied in these
last years~\cite{Teitelboim1,Carlip,Mann,Frolov}, in the vacuum case all
solutions are locally Minkowski (the Riemann tensor is zero); the extension
to the vacuum plus cosmological constant allows for the existence of the
rotating anti de Sitter regular black hole~\cite{Teitelboim1} (the scalar
curvature and the Ricci square invariants are constants proportional to $%
\Lambda$ and $\Lambda^2$.) The static (2+1)-charged black hole with
cosmological constant (the static charged BTZ solution) is singular (when $r$
goes to zero the curvature and the Ricci square invariants blow up).
Similarly as in the (3+1)-gravity, one may search for regular solutions in
(2+1)-gravity incorporating nonlinear electromagnetic fields to which one
imposes the weak energy conditions in order to have physically plausible
matter fields. One can look for regular solutions with nonlinear
electromagnetic fields of the Born-Infeld type~\cite
{Born,Garcia1,Garcia2,Gibbons,Fradkin,Deser} or electrodynamics of wider
spectra.

In this work, we are using electromagnetic Lagrangian $L(F)$ depending upon
a single invariant $F= 1/4 F^{a b} F_{ab}$, which we demand in the weak
field limit to be equal to the Maxwell Lagrangian $L(F) \longrightarrow - F/
4 \pi$, the corresponding energy momentum tensor has to fulfill the weak
energy conditions: for any timelike vector $u^a$, $u^a u_a= -1$ (we are
using signature -- + +) one requires $T_{a b} u^a u^b \geq 0$ and $q_a q^a
\leq 0$, where $q^a= T^a _b u^b$.

The action of the (2+1)-Einstein theory coupled with nonlinear
electrodynamics is given by
\begin{eqnarray}  \label{action}
S=\int \sqrt{-g} \left(\frac 1{16\pi} (R-2\Lambda) + L(F) \right) \,d^3x,
\end{eqnarray}
with the electromagnetic Lagrangian $L(F)$ unspecified explicitly
at this stage. We are using units in which $c=G=1$. The ambiguity
in the definition of the gravitational constant (there is not
Newtonian gravitational limit in (2+1)-dimensions) allows us to
maintain the factor $1/16 \pi$ in the action to keep the
parallelism with (3+1)-gravity. Varying this action with respect to
gravitational field gives the Einstein equations
\begin{eqnarray}
G_{ab} + \Lambda g_{a b}= 8 \pi T_{ab},
\end{eqnarray}
\begin{eqnarray}  \label{tensor electromagnetico}
T_{ab}= g_{ab} L(F)- F_{ac} F_{b}^{\,\,c} L_{_{,F}} ,
\end{eqnarray}
while the variation with respect to the electromagnetic potential $A_{a}$
entering in $F_{ab}= A_{b,a} - A_{a,b}$, yields the electromagnetic field
equations
\begin{eqnarray}  \label{poisson}
\nabla_{a} \left( F^{ab} L_{_{,F}} \right)=0,
\end{eqnarray}
where $L_{_{,F}}$ stands for the derivative of $L(F)$ with respect to $F$.

Concrete solutions to the dynamical equations above we present for the
static metric
\begin{eqnarray}  \label{metrica}
ds^{2}= - f(r) dt^{2} + \frac{dr^{2}}{f(r)} + r^{2} d \Omega^{2},
\end{eqnarray}
where $f(r)$ is an unknown function of the variable r. We restrict the
electric field to be
\begin{eqnarray}  \label{tensorr}
F_{a b} = E(r) \left ( \delta^{t }_{a } \delta^{r}_{ b} - \delta^{r }_{a}
\delta^{t}_{b} \right ).
\end{eqnarray}
The invariant $F$ then is given by
\begin{eqnarray}  \label{invariante}
2 F = -E^{2}(r),
\end{eqnarray}
thus the electric field can be expressed in term of the invariant $F$.
Substituting~(\ref{tensorr}) and~(\ref{invariante}) into the electromagnetic
field equations~(\ref{poisson}) we arrive at
\begin{eqnarray}  \label{Elec1}
E(r) L_{,F}= \frac{e}{r},
\end{eqnarray}
where $e$ is an integration constant. We choose $e= - q/4 \pi$ in order to
obtain the Maxwell limit. Then we have
\begin{eqnarray}  \label{Elec}
E(r) L_{,F}= - \frac{q}{4 \pi r}.
\end{eqnarray}
Using now~(\ref{invariante}) we express the derivative $L_{F}$ as function
of $r$, as follows
\begin{eqnarray}  \label{Elec2}
L_{,r}= \frac{q}{4 \pi r} E_{,r}.
\end{eqnarray}

We rewrite the Einstein's equations equivalently as
\begin{eqnarray}
R_{a b} = 8 \pi \left ( T_{a b} - T g_{ab} \right) + 2 \Lambda g_{a b}.
\end{eqnarray}
From~(\ref{tensor electromagnetico}) using~(\ref{tensorr}) and~(\ref
{invariante}) the trace becomes
\begin{eqnarray}
T= 3 L(F) + 2 E^{2}(r) L_{,F}.
\end{eqnarray}
As it was above pointed out, the Lagrangian $L(F)$ must satisfy: (i)
correspondence to Maxwell theory, i.e. $L(F) \longrightarrow - L/4 \pi$, and
(ii) the weak energy conditions: $T_{a b} u^a u^b \geq 0$ and $q_a q^a \leq 0
$, where $q^a= T^a _b u^b$ for any timelike vector $u^a$; in our case the
first inequality requires
\begin{eqnarray}
- (L + E^2 L_{,F}) \geq 0,
\end{eqnarray}
which can be stated equivalently as
\begin{eqnarray}  \label{condicion del lagrangiano}
L \leq E L_{,E} \longrightarrow L \leq \frac{q}{4 \pi r} \, E.
\end{eqnarray}
The norm of the energy flux $q_{a}$, occurs to be always less or equal to
zero; for $u^a$ along the time coordinate, $u^a= \delta^a_t/\sqrt{-g_{tt}}$,
one has the inequality $q_a q^a= - (L + L_{,F} E^2)^2 \leq 0$.

Assume now that one were taking into account additionally the scalar
magnetic field $B:=F_{\phi r}$, then the Maxwell equations would be
\begin{eqnarray}  \label{rit710}
{\frac{d}{dr}}[r E L_{,F}]=0,\,\,\,{\frac{d}{dr}} \frac{f}{r}B L_{,F}=0.
\end{eqnarray}
On the other hand, the Ricci tensor components, evaluated for the BTZ metric
(\ref{metrica}), would yield the following relation
\begin{eqnarray}  \label{rit11}
A := R_{tt} +f^2 R_{rr}=0,
\end{eqnarray}
while the evaluation of the same relation using the electromagnetic
energy-momentum would give
\begin{eqnarray}
A=-8 \pi L_{,F}(\frac{f}{r}B)^2.
\end{eqnarray}
Therefore, the scalar magnetic field should be equated to zero , $B=0$ ,
thus the only case to be treated is just the one with the electric field $E$.

As far as the Einstein equations are concerned, the $R_{tt} ( =- f^2 R_{_{r
r}}) $ and $R_{_{\Omega \Omega}}$ components yield respectively the
equations
\begin{eqnarray}  \label{tt}
f_{,rr}+ \frac{f_{,r}}{r} = -4\Lambda + 16 \pi \left ( 2 L(F) + E^{2} L_{,F}
\right),
\end{eqnarray}
\begin{eqnarray}  \label{omegaomega}
f_{,r}= -2 \Lambda r + 16 \pi r \left (L(F) + E^{2} L_{,F} \right).
\end{eqnarray}
If one replaces $f_{,r}$ from (\ref{omegaomega}) and its derivative
$f_{,rr}$ into (\ref{tt}) one arrives, taking into account the
equation (\ref{Elec2}), at an identity. Therefore one can forget
the equation (\ref{tt}) and integrate the relevant Einstein
equation (\ref{omegaomega}):
\begin{eqnarray}  \label{solucion penultima}
f(r)= - M - \Lambda r^{2}  \nonumber \\
+ 16 \pi \int r \left [ L(F(r)) + E^2 L_{,F} \right] dr.
\end{eqnarray}

Summarizing we have obtained a wide class of solutions, depending on a
Lagrangian $L(E)$, given by: \\the metric
\begin{eqnarray}
ds^{2}= - f(r) dt^{2} + \frac{dr^{2}}{f(r)} + r^{2} d \Omega^{2},
\end{eqnarray}
the structural function
\begin{eqnarray}  \label{solucion ultima}
f(r)= - M - (\Lambda - 2 C) r^{2}  \nonumber \\
+ 4 q \int \left [r \int \frac{E_{,r}}{r} dr - E \right] dr,
\end{eqnarray}
which is obtained from~(\ref{solucion penultima}) by using~(\ref{Elec2})
and~(\ref{invariante}), where $C$ is a constant of integration, \\and the
Lagrangian $L(E)$ is constrained to
\begin{eqnarray}
L_{,r}= \frac{q}{4 \pi r} E_{,r},
\end{eqnarray}
We recall that the Lagrangian and the energy momentum tensor have to fulfill
the conditions quoted above.

We present now various particular examples: \\The static charged BTZ
solution~\cite{Teitelboim1} is characterized by the function
\begin{eqnarray}
-g_{tt}= f = - M + \frac{r^{2}}{l^{2}} - 2 q^{2} ln \, r,  \label{BTZ}
\end{eqnarray}
the Lagrangian and the electric field
\begin{eqnarray}
L(E)= \frac{1}{8 \pi}E^2 = \frac{1}{8 \pi} \frac{q^2}{r^2}, \,\,\,\,\, E(r)=%
\frac{q}{r},
\end{eqnarray}
where $C$ has been equated to zero and $\Lambda= -1/l^{2}$. It is worthwhile
to point out that the static charged BTZ black hole is singular at $r=0$.

Other interesting example arises in the Born-Infeld electrodynamics --
nonlinear charged (2+1)--black-hole~\cite{nuestro}. In this case the
structural function is
\begin{eqnarray}
-g_{tt}= f = - M - (\Lambda - 2 b^{2}) r^{2} -2 b^{2}r \sqrt{r^{2}+
q^{2}/b^{2}}  \nonumber \\
- 2 q^{2} ln (r +\sqrt{r^{2}+q^{2}/b^{2}} ),
\end{eqnarray}
and the Lagrangian and the electric field are given by
\begin{eqnarray}
L(F)=-\frac{b^{2}}{4 \pi} \left (\sqrt{1+2 \frac{F}{b^{2}}}-1 \right)=
\nonumber \\
-\frac{b^2}{4 \pi} \left( \frac{r}{\sqrt{r^2 + q^2/b^2}} -1 \right),
\nonumber \\
E(r)= \frac{q}{\sqrt{r^2 + q^2/b^2}}, \hspace{2 cm}
\end{eqnarray}
where $b$ is the Born-Infeld parameter, and $C= b^2$. This solution fulfills
the weak energy conditions and it is singular at $r=0$. From the Ricci and
Kretschmann scalars it follows that in this case there is a curvature
singularity at $r=0$~\cite{nuestro}.

A new class of solution, which is regular everywhere, is given by the
structural function of the form
\begin{eqnarray}
f(r)=- M - \Lambda r^2 - q^2 ln(r^2+a^2)
\end{eqnarray}
where $M$, $a$, $q$ and $\Lambda$ are free parameters. The Lagrangian and
the electric field are given by
\begin{eqnarray}
L(r)= \frac{q^2}{8 \pi} \, \frac{(r^2 - a^2)}{(r^2 + a^2)^2},  \nonumber \\
E(r)= \frac{q r^3}{(r^2+a^2)^2}.
\end{eqnarray}
This Lagrangian requires to set $C=0$. The Lagrangian and the electric field
satisfy the weak energy conditions~(\ref{condicion
del lagrangiano}). To
express the Lagrangian in terms of $F$ or equivalently $E$, one has to write
$r$ in terms of $E$ by solving the quartic equation for $r(E)$, this will
give rise an explicit $r$ containing radicals of $E$, which introduced in $%
L(r)$, finally will bring $L$ as function of $E$. The expression $L(E)$ is
not quite illuminating, thus we omit it here.

To establish that this solution is regular one has to evaluate the curvature
invariants~\cite{Wald}. The non-vanishing curvature components, which occur
to be regular at $r=0$, are given by:
\begin{eqnarray}
R_{0 1 1 0}= \frac{q^{2}(a^2- r^2)}{(r^2+a^2)^2} + \Lambda,
\end{eqnarray}
\begin{eqnarray}
R_{0 2 0 2}= - f(r) \left ( \frac{q^{2} r^2}{ r^{2}+ a^{2}} + \Lambda r^2
\right),
\end{eqnarray}
\begin{eqnarray}
R_{1 2 1 2}= f(r)^{-1} \left ( \frac{q^{2} r^2}{ r^{2}+ a^{2}} + \Lambda r^2
\right),
\end{eqnarray}
where 0,1,2 stand respectively for t,r and $\Omega$.

Evaluating the invariants $R$, and $R_{ab}R^{ab}$ one has
\begin{eqnarray}
R&=&\frac{2q^2(r^2+3a^2)}{(r^2+a^2)^2}+6\Lambda  \label{escalarR}
\\
R_{ab}R^{ab}&=&12\Lambda^2+4q^4\frac{r^4+2r^2a^2+3a^4}{(r^2+a^2)^4}
\nonumber  \\
&&+\frac{8\Lambda q^2(3a^2+r^2)}{(r^2+a^2)^2}.
\label{escalarRabRab}
\end{eqnarray}
Since the metric, the electric field and these invariants behave
regularly for all values of $r$, we conclude that this solution is
curvature regular everywhere. Nevertheless, for solutions without
any horizon or black hole solutions with an inner and outer
horizons, at $r=0$ a conical singularity may arise.\\ At $r=0$ the
function $f(r)$ becomes $f(0)=-M-q^2\ln (a^2)$.\\ Thus for $M$
positive, $M>0$, and $a$ in the range $0<a<1$, the value of $f(0)$
will be $f(0)=-M+q^2\ln (1/a)^2$ , which will be positive, say
$f(0):=\beta ^2$, if $\ln (1/a)^2>M/q^2$. In such a case, for
$0<\beta <1$ the solutions will show angular deficit since the
angular variable $\Omega $, which originally runs
$0\leq\Omega<2\pi$ will now run $0\leq\Omega <2\beta \pi$; the
parameter $a$ can be expressed in terms of $\beta $, $q$ and $M$ as
$a^2=\exp [-(\beta^2+M)/q^2]$. For $\beta =1$, there will be no
angular deficit, the ratio of the perimeter of a small circle
around $r=0$ to its radius, as this last tends to zero, will be
$2\pi $.\\ If one allows $M$ to be negative, $M<0$, and $a$ to take
values in the interval $0<a<1$, then $f(0) $ will be always
positive, in this case one can adopt the following parametrization:
$-M=\beta^2\cos^2\alpha$, $q^2\ln(1/a)^2=\beta^2\sin^2\alpha$,
therefore $f(0)=\beta ^2$. One will have angular deficit if
$0<\beta <1$ , and for $\beta =1$ the resulting (2+1) space-time
will be free of singularities. Another possibility with positive
$f(0)=\beta ^2$ arises for $M<0$, and $a>1$, $f(0)$ can be
parameterized as $-M=\beta ^2\cosh^2\alpha$,
$q^2\ln(1/a)^2=\beta^2\sinh^2\alpha$. Again the values taken by
$\beta $ will govern the existence of angular deficit, for $\beta
=1$ the solutions will be regular.\\ If $f(0)$ is negative,
$f(0)=:-\beta^2$, the character of the coordinates $t$ and $r$
changes, the coordinate $t$ becomes space--like, while $r$ is now
time--like and one could think of the singularities, if any, as
causal structure singularities because they could arise at the
``time'' $r=0$.\\ In what follows we shall treat the parameter $a$
as a free one, having in mind the above restrictions to have
solutions free of conical singularities.

To establish that this solution represents a black hole, one has to
demonstrate the existence of horizons, which require the vanishing of the $%
g_{tt}$ component, i.e., $f(r)=0$. The roots of this equation give the
location of the horizons (inner and outer in our case). The roots -- at most
four -- of the equation $f(r)=0$ can be expressed in terms of the Lambert $%
W(r)$ function
\begin{eqnarray}
r_{1,2,3,4}= \pm [ exp(\frac{\Lambda a^2-M}{q^2} -  \nonumber \\
LW[\frac{\Lambda}{q^2} exp \left ( \frac{\Lambda a^2-M}{q^2} \right)]) - a^2
]^{\frac{1}{2}}.
\end{eqnarray}
There arise various cases which depend upon the values of the parameters:
four real roots (two positive and two negative roots, the negative roots
have to be ignored), two complex and two real roots, two complex and one
real positive root (the extreme case), and four complex roots ( no black
holes solutions.) Although this analytical expression for the Lambert
function can be used in all calculations, (we recall that Lambert function
fulfills the following equation $ln (LW(x))+ LW(x)= ln(x)$), it occurs also
useful to extract information from the graphical behavior of the our $f(r)$
(see figures).

Analytically one can completely treat the extreme black hole case; for it,
the derivative of $f(r)$ has to be zero, $\partial_{r}(f(r))=0$ , at the $%
r_{extr}$, this gives
\begin{eqnarray}  \label{rextr}
r_{extr}= \sqrt{-a^2 - \frac{q^2}{\Lambda}} > 0
\end{eqnarray}
for $\Lambda < 0$. From this expression one concludes that the following
inequality holds: $a^2 < - q^2/ \Lambda$. Entering now $r_{extr}$ into $%
f(r)=0$ one obtains a relation between the parameters involved, which can be
solved explicitly for the mass--the extreme one--
\begin{eqnarray}
M_{extr}= a^{2} \Lambda + q^2 \left( 1 + ln \left[ \frac{- \Lambda}{q^2}
\right ] \right),
\end{eqnarray}
this $M_{extr}$ varies its values depending on the values given to the
parameters $a$, $q$ and $\Lambda$. We have an extreme black hole
characterized by negative cosmological constant, $\Lambda < 0$, and positive
extreme mass, $M_{extr} > 0$, if the parameter $a$ is restricted by the
inequality $a^{2} < - (q^2(1 + ln(- \Lambda/q^2)))/ \Lambda$.

For other values of the mass $M$, one distinguishes the following
branches: if $M >M_{extr}$ one has a black hole solution, and if $M
< M_{extr}$ there are no horizons.

In FIG. 1 we draw the graph of $f(r)$ which corresponds to regular
solutions for a fixed values of $M$ and changing the values of the
parameters $%
\Lambda $, $q$. In FIG. 2 we draw the graph of $f(r)$ corresponding to
solutions which exhibit a conical singularity at $r=0$, for  $f(0)=1/2$,
keeping $M$ fixed while $\Lambda $ and $q$ change.

\begin{figure}[ht]
\centerline{ \psfig{file=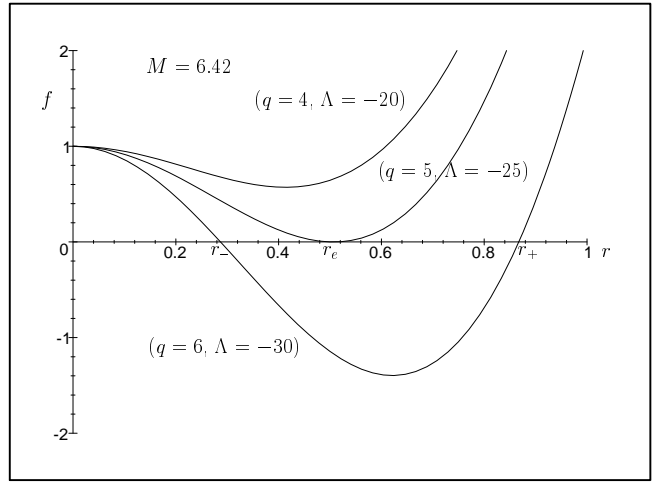,width=9cm} }
\caption{Behavior of $-g_{tt}$ for $M=6.421$ and for
different values of $q$ and $\Lambda$ corresponding to regular
black hole ($q=6$, $\Lambda=-30$), regular extreme black hole
($q=5$, $\Lambda=-25$), horizon-free regular solution ($q=4$,
$\Lambda=-20$), where $r_{-}=0.28$ is the inner horizon,
$r_{e}=0.50$ is the extreme horizon, and $r_{+}=0.86$ is the event
horizon.}
\label{fig:A}
\end{figure}
\begin{figure}[ht]
\centerline{ \psfig{file=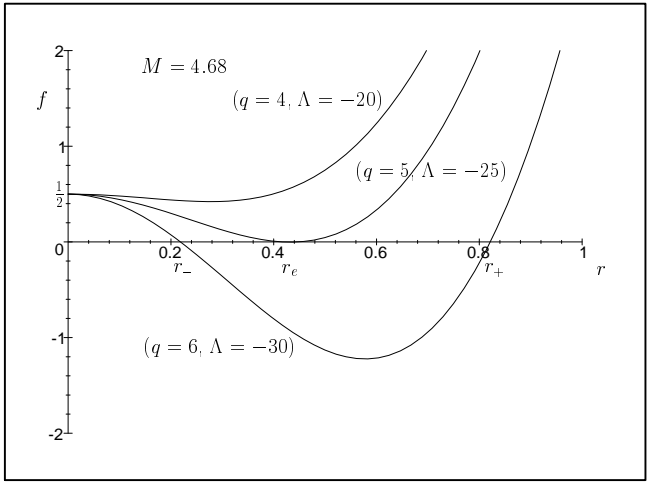,width=9cm} }
\caption{Behavior of $-g_{tt}$ for $M=4.68$ and for
different values of $q$ and $\Lambda$ for solutions with angular
deficit $0 \leq \Omega < \pi /
\sqrt{2} 
$ corresponding to conical singular black hole ($q=6$,
$\Lambda=-30$), conical singular extreme black hole ($q=5$,
$\Lambda=-25$), and conical naked singular solution ($q=4$,
$\Lambda=-20$), where $r_{-}=0.21$ is the inner horizon,
$r_{e}=0.43$ is the extreme horizon, and $r_{+}=0.82$ is the event
horizon.} \label{fig:B}
\end{figure}


If one were interested in the thermodynamics of the obtained solution one
would to evaluate the temperature of the black hole, which is given in terms
of its surface gravity by~\cite{Visser,Brown}
\begin{eqnarray}  \label{T}
k_{_{B}} T_{_{H}} = \frac{\hbar}{2 \pi} \, k.
\end{eqnarray}
In general, for a spherically symmetric (and for circularly symmetric in
(2+1)-dimensions) system the surface gravity can be computed via (for our
signature)
\begin{eqnarray}  \label{k}
k=- \lim_{r \rightarrow r_{_{+}}} \left [\frac{1}{2} \frac{\partial_{r}
g_{tt}}{\sqrt{- g_{tt} g_{rr}}} \right ],
\end{eqnarray}
where $r_{_{+}}$ is the outermost horizon. For our solution we have from~(%
\ref{solucion ultima}),~(\ref{T}) and~(\ref{k}) that
\begin{eqnarray}  \label{temp}
k_{_{B}} T = \frac{\hbar}{2 \pi} \left(- \Lambda r_{_{+}} - \frac{q^2
r_{_{+}}}{r^2_{_{+}} + a^2} \right ).
\end{eqnarray}
Since in our case there is no an analytical expression of
$r_{_{+}}$ in terms of elementary functions, one can not give a
parameter dependent expression of~(\ref{temp}). It is easy to check
that when $q=0$, $T$ in~(\ref {temp}) reduces to the BTZ
temperature. In the extreme case~(\ref{rextr}), the temperature
vanishes in~(\ref{temp}). The entropy can be trivially obtained
using the entropy formula $S= 4 \pi r_{_{+}}$. Other thermodynamic
quantities such as heat capacity and chemical potential can be
computed as in~\cite{Brown}.

To achieve the maximal extension of our regular black solutions one has to
follow step by step the procedure presented in~\cite{Wald} determining first
the Kruskal-Szekeres coordinates, and to proceed further to draw the Penrose
diagrams.

Informative discussions with Jorge Zanelli, Ricardo Troncoso,
Rodrigo Aros, and Eloy Ay\'{o}n-Beato are gratefully acknowledged.
This work was supported in part by FONDECYT-Chile 1990601,
Direcci\'{o}n de Promoci\'{o}n y Desarrollo de la Universidad del
B\'{\i}o-B\'{\i}o through Grant No 983105-1 (M.C.), FONDECYT-Chile
1980891, CONACYT-M\'exico 3692P-E9607, 32138E (A.G.) and in part
by Dicyt de la Universidad de Santiago de Chile (M.C., A.G.).

\end{document}